\documentclass[aps,prl,twocolumn,showpacs,superscriptaddress]{revtex4}   % for review and submission with two columns
\usepackage{hyperref}  % needed for live references
\usepackage{graphicx}  % needed for figures
\usepackage{dcolumn}   % needed for some tables
\usepackage{bm}        % for math
\usepackage{amssymb}   % for math
\usepackage{verbatim}  % for comment
\usepackage{epstopdf}
\usepackage{color}
\usepackage{soul}
\usepackage{amsmath}
\usepackage{amsfonts}

\newcommand{\bea}{\begin{eqnarray}}
\newcommand{\eea}{\end{eqnarray}}
\newcommand{\beq}{\begin{equation}}
\newcommand{\eeq}{\end{equation}}

\newcommand{\eq}[1]{(\ref{#1})}

\newcommand{\br}{{\bf r}}

\newcommand{\bE}{{\bf E}}

\newcommand{\bd}{{\bf d}}

\newcommand{\cbE}{\boldsymbol{\mathbf{\cal E}}}
\newcommand{\dip}{{\cal D}}
\newcommand{\G}{{\rm G}}

\begin{document}

\title{Shifts of a resonance line in a dense atomic sample}

\author{Juha Javanainen}
\affiliation{Department of Physics, University of Connecticut, Storrs, Connecticut 06269-3046}
\author{Janne Ruostekoski}
\affiliation{Mathematical Sciences, University of Southampton,
Southampton, SO17 1BJ, UK}
\author{Yi Li}
\affiliation{Department of Physics, University of Connecticut, Storrs, Connecticut 06269-3046}
\author{Sung-Mi Yoo}
\affiliation{Department of Physics, University of Connecticut, Storrs, Connecticut 06269-3046}
\begin{abstract}
We study the collective response of a dense atomic sample to light essentially exactly using classical-electrodynamics simulations. In a homogeneously broadened atomic sample there is no overt Lorentz-Lorenz local field shift of the resonance, nor a collective Lamb shift. However, addition of inhomogeneous broadening restores the usual mean-field phenomenology.
\end{abstract}
\pacs{42.50.Nn,32.70.Jz,42.25.Bs}
\maketitle
Textbook arguments~\cite{JAC99,BOR99} tell us that in a dielectric medium the local electric field $E_l$ seen by an atom (molecule) is different from the macroscopic electric field $E$ by an amount proportional to the polarization $P$ of~the medium, $E_{l} = E +{P}/{3\epsilon_0}$. This is the origin of the local-field corrections in electrodynamics embodied in the Clausius-Mossotti and Lorentz-Lorenz relations. As a result, the frequency dependence of the microscopic polarizability and the macroscopic susceptibility are different. If the polarizability has a Lorentzian line shape then so does the susceptibility, but the resonance is shifted by what is known as the Lorentz-Lorenz (LL) shift~\cite{LOR11}. The LL shift serves as the generic frequency scale for other density dependent phenomena in an atomic sample such as collisional self-broadening of absorption lines~\cite{LEW80,WEL11} and collective Lamb shift (CLS)~\cite{FRI73,SVI10,KEA12,ROH10,SCU10}.

Local-field corrections are a standard workhorse in solid and liquid media. On the other hand, in a resonant atomic gas a density conducive to LL shift and CLS results in an optically thick sample, which might explain the sparsity of laser spectroscopy era experiments. There are careful experiments on related phenomenology that  agree with the respective theory~\cite{MAK91,WAN97,VAN99,ROH10,KEA12}, but except for the nuclear-physics experiment of Ref.~\cite{ROH10} the published experiments we know of deal with inhomogeneously broadened samples with a substantial line broadening due to the motion of the atoms.	{ Atomic-physics experiments with cold and dense clouds such as those in Ref.~\cite{BOU13} are presently underway~\cite{BRO13}}. Optically thick samples are needed for a good quantum interface between photons and matter~\cite{HAM10}, so that local-field effects, and more generally, cooperative response of matter to light, are likely to become issues in the quest toward quantum technologies.

Here we study the cooperative response of a dense atomic sample to light essentially exactly~\cite{JAV97} using classical-electrodynamics simulations~\cite{JAV99LQD,PIN04,AKK08,CHO12,Jenkinsoptlattice,JEN12a,JEN12INH,Castin2013} in a slab geometry, analogously to  theory~\cite{FRI73} and experiments~\cite{KEA12} on CLS. A homogeneously broadened sample with fixed atomic positions does not exhibit a simple line shift, but instead subtle modifications of the absorption and fluorescence spectra. However, when we add inhomogeneous broadening~\cite{JEN12INH} to the atomic samples, the traditional phenomenology of local-field corrections together with density-dependent collective effects reemerges. Basically, in a homogeneously broadened sample the correlations between nearby atoms established by the dipole-dipole interactions are important, while inhomogeneous broadening suppresses the correlations and makes the sample behave more like a continuous polarization.

Let us first look at the logical status of the LL shift and CLS as in Ref.~\cite{FRI73} from the standpoint of our earlier analysis~\cite{JAV97,RUO97}. Briefly, we derived from quantum field theory of light and matter coupled equations of motion for the correlation functions of atomic density and polarization. In the limit of low light intensity, these involve atomic position correlation functions $\rho_p(\br_1,\ldots,\br_p)$ and correlation functions with  density at points $\br_1,\ldots,\br_{p-1}$ and polarization at $\br_p$, ${\bf P}_p(\br_1,\ldots,\br_{p-1};\br_p)$. For a $J=0\rightarrow J'=1$ atomic transition the response of the medium is isotropic, and we have a hierarchy of equations of motion for the correlation functions
\begin{align}
&\dot{\bf P}_p(\br_1,\ldots,\br_{p-1};\br_p) =\nonumber\\
&(i\Delta-\gamma){\bf P}_p(\br_1,\ldots,\br_{p-1};\br_p)
+i\zeta{\cbE}_0(\br_p)\rho_k(\br_1,\ldots,\br_p )\nonumber\\
&+i\zeta\sum_{q\ne p}\G(\br_p-\br_q){\bf P}_p(\br_1,\ldots,\br_{q-1},\br_{q+1}\dots,\br_p;\br_q)\nonumber\\
&+i\zeta\int d^3r_{p+1}\G(\br_p-\br_{p+1}){\bf P}_{p+1}(\br_1,\ldots,\br_p;\br_{p+1}),
\label{EXACT}
\end{align}
with $p=1,2,\ldots$.
This is a near-monochromatic version of the theory, with ${\bf P}_p$ being positive-frequency parts of quantities oscillating at the ``laser frequency'' $\omega$. $\Delta=\omega-\omega_0$ is the detuning from the atomic resonance $\omega_0$, $\gamma$ is the HWHM linewidth of the transition, {$\zeta=\dip^2/\hbar$,} $\dip$ is the dipole moment matrix element, and {$\epsilon_0 {\cbE}_0$} would be the electric displacement of the driving light if the matter were absent. $\G$ is the dipole field propagator, a $3\times3$ matrix such that $\G(\br-\br') \bd$ is the usual~\cite{JAC99} electric field at $\br$ from a dipole $\bd$ at $\br'$.

The peculiar feature of Eq.~\eq{EXACT} is the third line. It describes $p$ atoms interacting with each other via the dipole-dipole interaction. Suppose we factorize the lowest nontrivial correlation function ${\bf P}_2(\br_1;\br_2)$ as ${\bf P}_2(\br_1;\br_2)=\rho(\br_1){\bf P}(\br_2)$. This assumption says that there are no correlations between the positions and dipole moments of the atoms, in violation of the fact that the dipole-dipole interactions depend on the positions. We then have an equation for atom density and polarization,
\bea
\dot{\bf P}(\br) &=& (i\Delta-\gamma){\bf P}(\br)+i\zeta\rho(\br){\cbE}_0(\br)\nonumber\\
&+&i\zeta\rho(\br)\int d^3r\, \G(\br-\br'){\bf P}(\br')\,.
\label{APPR}
\eea
In this mean-field approximation polarization is taken to be a continuous field. As this approach no longer retains the information about the precise positions of the atoms, correlations induced by scattered photons between nearby atoms that depend on the spatial distribution of the atoms are generally also lost. The model \eq{APPR} reproduces the standard electrodynamics of a continuous polarizable medium~\cite{BOR99}.
The integral is not absolutely convergent. As has been discussed before~\cite{BOR99,FRI73,JAV97}, one in effect carries it out as if $\G(\br-\br')$ were not singular, and adds to $\G$ an extra diagonal term $\delta(\br-\br')/3\epsilon_0$ to account for the singularity. This is where the local-field corrections and the LL shift enter.

For a slab configuration, a uniform-density medium restricted to the interval $z\in[0,h]$ and a plane wave with the wave number $k=\omega/c$ propagating in the $z$ direction, the stationary solution to Eq.~\eq{APPR} may be found exactly, including the field transmitted through the slab. In the limit of asymptotically small density $\rho$ of the medium, the absorption line is Lorentzian and is shifted by
\beq
\Delta_L = \Delta_{LL}\! -\!  \frac{3}{4}\Delta_{LL}\left(1\! -\! \frac{\sin 2hk}{2hk} \right);\quad \label{COLLAMB}
\Delta_{LL} = - \frac{\rho\dip^2}{3\epsilon_0\hbar}
\eeq
from the atomic resonance. Here $\Delta_{LL}$, a red shift, is the standard LL shift, and $\Delta_L$ is the CLS as in Ref.~\cite{FRI73}. In the present formulation the CLS is a combination of the LL shift and the etalon effect because of the reflections of light from the front and back surfaces of the sample. This is the CLS verified in the experiments~\cite{KEA12}.

It is obvious from Ref.~\cite{FRI73} that the derivation of the CLS $\Delta_L$ also came down to what we term mean-field theory, and the same applies to many other analyses such as in Ref.~\cite{SVI10}. This approximation ignores the correlations between nearby atoms that might arise from dipole-dipole interactions, and as such is uncontrolled. This is why we choose to solve the steady state of Eqs.~\eq{EXACT} essentially exactly~\cite{JAV99LQD} using classical-electrodynamics simulations~\cite{JAV99LQD,PIN04,AKK08,CHO12,Jenkinsoptlattice,JEN12a,JEN12INH,Castin2013}.

We characterize the incident monochromatic field driving the atomic dipoles with the complex amplitude $\cbE_0(\br)$ and likewise all other quantities oscillating at the frequency $\omega$. We have the $N$ atoms fixed at positions $\br_i$, $i=1,\ldots,N$, each with an assumedly isotropic polarizability $\alpha$. In addition to the incident field, each atom $i$ at position $\br_i$ is illuminated by scattered radiation from all the $N-1$ other atoms, $\bE^{(j)}_S(\br)$ $(j\neq i)$, so that the total external field driving the atom is $\bE_{\rm}(\br_i) = \cbE_0(\br_i)+\sum_{j\ne i} \bE^{(j)}_S(\br_i)$. This induces the dipole moment $\bd_i\equiv\bd(\br_i)=\alpha \bE(\br_i)$, which will in its turn emit the electric field $\bE^{(i)}_S(\br)= \G(\br-\br_i)[\alpha\bE(\br_i)]$.
We find a closed set of linear equations for the amplitudes $\bE({\br_i})$,
\beq
\bE({\br_i}) = \cbE_0({\br_i}) + \alpha\sum_{j\ne i} \G(\br_i-\br_j)\bE(\br_j)\,.
\label{LINEQS}
\eeq
Having solved it numerically, we have the electric field amplitude everywhere in the form
\beq
\bE(\br) = \cbE_0(\br) + \alpha\sum_{i}\G(\br-\br_i)\bE(\br_i)\,.
\eeq

The polarizability of the quintessential two-level atom is $\alpha=-\dip^2/[\hbar(\Delta+i\gamma)]$, with $\gamma=\dip^2k^3/6\pi\hbar\epsilon_0$. A two-level atom has a preferred direction for polarization, but with our tacit assumption that we are dealing with the $J=0\rightarrow J'=1$ transition the dipole of an atom and the polarization of the driving light are actually parallel.  There will be adjustments for other level schemes even in the limit of low light intensity, let alone when optical pumping is a factor, but we do not go into the details. The LL shift turns out to be $\Delta_{LL} = -2\pi\gamma \rho k^{-3} $. The abundance of powers of the wave number $k$ in our formulas reflects the fact that the natural unit of length for optical response is $k^{-1}$.

In our numerical experiments we study a circular disk with radius $R=\sqrt{256/\pi}\,k^{-1}$, so that the area is $A=256\,k^{-2}$. We vary the  thickness of the disk $h$ but keep the density $\rho=N/hA=2 k^3$ constant, so the number of atoms $N$ varies accordingly. A circularly polarized plane wave comes in perpendicular to the face of the disk. Analogous numerical experiments, although for different purposes, have been described in Refs.~\cite{Jenkinsoptlattice,CHO12}. In this context absorption means destructive interference of the incoming light and the light radiated by the atoms of the disk in the forward direction. We denote the fractional reduction of the energy density of light by $1-T$, where $T$ is the coefficient of transmission.  We adapt the method to calculate the absorption coefficient from Ref.~\cite{CHO12}. Occasionally we also  compute the back-scattered power by integrating the radial component of the Poynting vector of the scattered radiation over a large-radius hemisphere that covers the disk on the side of the incoming light. Even if local-field effects are a major theme here, we always analyze observable quantities outside of the sample and thus avoid the question of the operational meaning of the fields inside the sample.

The overall protocol is that we generate a number of random samples, from 64 to millions, of atomic positions evenly distributed inside the disk, compute the absorption as a function of the detuning $\Delta$ for each sample, and average the results. At times we also compute the dependence of back-scattered power on frequency. By energy conservation, for an infinite radius of the disk the line shapes in absorption and back-scattering should be the same. A comparison strongly suggests that our observations are not an artifact of the rather small radius of the disk. We express the final results in terms of optical thickness (depth, density) $D$ defined as $D=-\ln T$. The advantage is that in a medium that obeys Beer's law the line shape of optical thickness $D$ would be independent of  the thickness $h$ of the sample.

The numerical experiments are similar to the real experiments of Keaveney et al.~\cite{KEA12}, with the significant exception that they had thermal samples at temperatures substantially higher than the room temperature while our atoms are standing still. Our simulations also differ from the experiments in that the densities are lower. This is because the computer time for a simulation grows as the cube of the atom number, and our runs add up to  $\sim10^5$ hours of CPU time as is.

\begin{figure}[t]
\includegraphics[width=8.5cm]{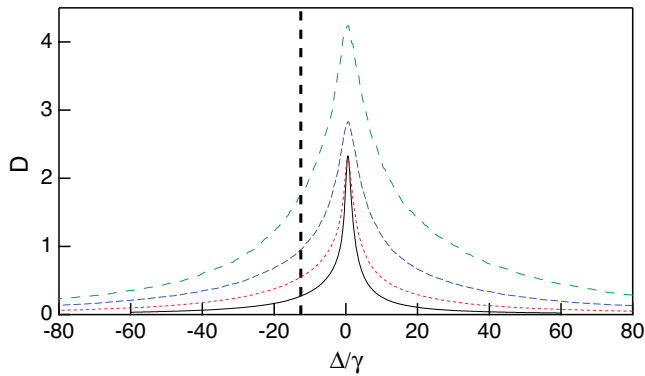}
\caption{(Color online) Optical depth $D$ versus detuning $\Delta$ in a homogeneously broadened sample for sample thicknesses $hk=0.25$, 0.5, 1.0 and 2.0, from bottom to top; the corresponding atom numbers are $N = 128$, 256, 512 and 1024. The dashed vertical line shows where the center of the line would be if the naive Lorentz-Lorenz shift applied.}
\label{HOMLINE}
\end{figure}

Figure~\ref{HOMLINE} shows the optical thickness $D$ as a function of detuning $\Delta$ for the sample thicknesses $hk=0.25$, $0.5$, $1.0$ and $2.0$, with the corresponding atom numbers $N=128$, $256$, $512$, and $1024$.  For comparison we also give the predicted  LL shift for this atom density as the dashed vertical line. The absorption lines are not Lorentzian. While the line broadens with increasing atom number and may be noticeably asymmetric, the maximum moves very little. The shift, if any, is at most a few percent of the LL shift. There is no manifest LL shift, nor a CLS.

The traditional density dependent shifts are predicted from mean-field theory that ignores the correlations between the dipoles. Here all correlations are included, and there is no mystery to the observation that our results differ from the established predictions. This, however, leaves the question of why experiments~\cite{KEA12} that by definition include all orders of dipole-dipole correlations agree with theoretical arguments~\cite{FRI73} that do not.

In real experiments with gaseous media the environment of a radiating atom is complex. The atom moves, there are atom-atom collisions, and the atoms collide with the walls of the container. Overall, the electric field that each atom sees changes as a function of time because both the spectator atom and the other atoms move. In the zeroth order picture of laser spectroscopy all of this is represented by inhomogeneous broadening: In the laboratory frame the resonance frequency of an atom depends on its velocity because of the Doppler shift, and accordingly, the resonance frequencies of the atoms are simply regarded as random quantities. Here we adopt this generic model.
\begin{figure}[t]
\includegraphics[width=0.53\columnwidth]{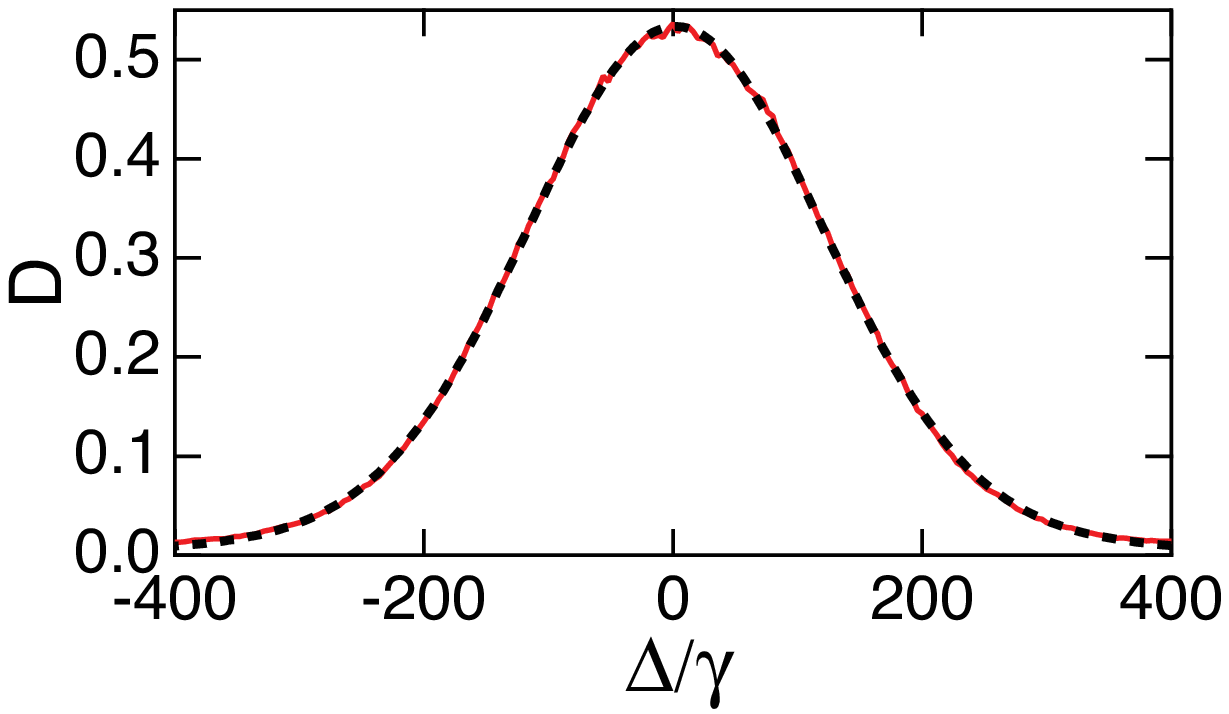}
\includegraphics[width=0.45\columnwidth]{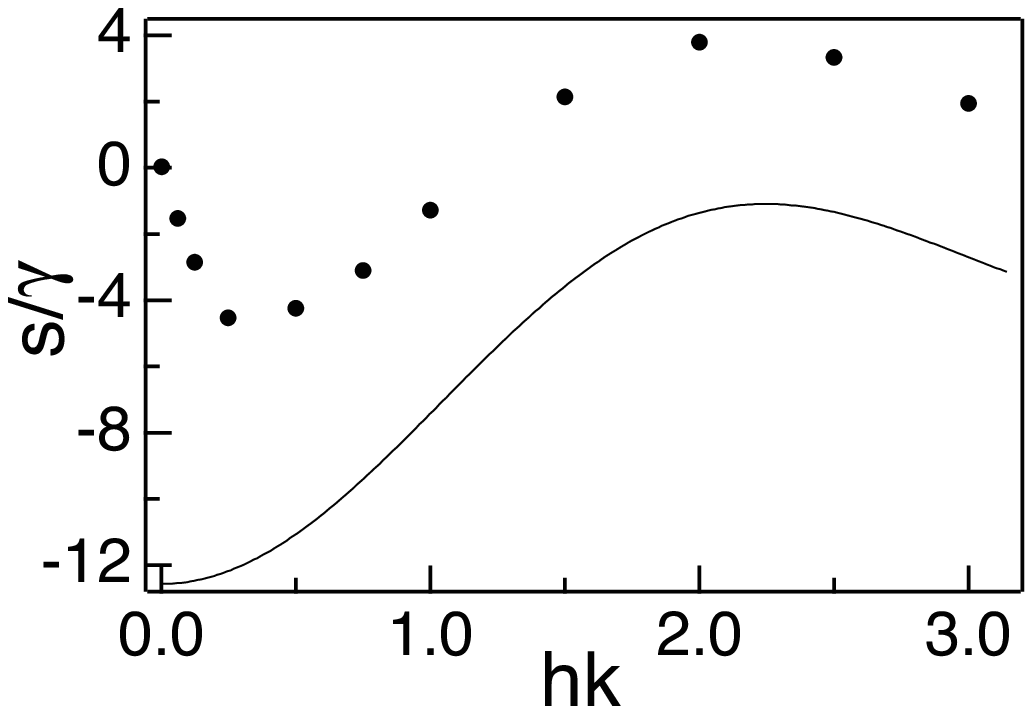}
\caption{(Color online) Left: Optical depth of the sample $D$ with thickness $hk=1.5$, and hence $N=768$ atoms, as a function of the detuning for a sample with the inhomogeneous linewidth $\omega_D=100\,\gamma$. This numerical experiment (solid red line) is an average of 1024 samples, the fit with a Voigt profile (dashed black line) has the parameters $s=2.15\,\gamma$, $\Gamma=17.74\,\gamma$ and $\Omega_D=112.83\,\gamma$.
Right: The shift of the absorption line $s$ plotted as a function of the thickness of the sample $h$ as solid circles.  The statistical error bars are smaller than the size of the circles. Also shown as a solid line is the collective Lamb shift, Eq.~\eq{COLLAMB}.}
\label{INHLINE}
\end{figure}

We repeat the numerical experiments with the atoms in the circular disk, except that this time we assume that the resonance frequency of each atom is also shifted by a Gaussian random variable with zero mean and the rms value $\omega_D= 100\,\gamma$. This value would be a reasonable estimate for the D lines in a room-temperature alkali vapor. An example spectrum is shown in Fig.~\ref{INHLINE}, left panel. The line shape has the appearance common in the spectroscopy of inhomogeneously broadened samples. Accordingly, we fit it with the Voigt profile $V(\Delta;\Gamma,\Omega_D)$, convolution of a Lorentzian with the HWHM width $\Gamma$ and a Gaussian with the rms width $\Omega_D$. More precisely, the fit function is $M\, V(\Delta-s;\Gamma,\Omega_D)$, where the fit parameters are the overall scale $M$, the shift of the resonance $s$, the linewidth $\Gamma$, and the inhomogeneous broadening $\Omega_D$. The quality of the fit is very good.

We plot the shift of the resonance $s$ as a function of the thickness of the disk $h$ in Fig.~\ref{INHLINE}, right panel, as filled circles. The shift is a small fraction of the total width of the spectrum but it is highly reproducible; the statistical $1\sigma$ error bars fit inside the circles.  The shift tends to zero at small thicknesses. Obviously, the physics becomes two-dimensional in this limit, three-dimensional density is meaningless, and  eventually there are too few atoms to influence one another anymore. We also plot the CLS, Eq.~\eq{COLLAMB}, as a solid line. Numerical data and theory show similar oscillations, albeit differing approximately by an additive constant. There was an additive term fitted to the experiments~\cite{KEA12}, too, before they gave an agreement with Eq.~\eq{COLLAMB}. The agreement of our numerical experiments with the theory is on a similar footing as in the real experiments.

For the $J=0\rightarrow J'=1$ transition the self-broadening of the atomic line due to the dipole-dipole interactions, part of the collisional interactions between the atoms that adds to the natural linewidth, is predicted to be $\gamma'\simeq2\pi\,\sqrt{3}\,\rho k^{-3}\gamma\simeq 22\,\gamma$~\cite{LEW80}. Optical experiments in dense samples~\cite{MAK91,WEL11,KEA12} have shown good agreement with theory~\cite{LEW80}. Our fitted value $\Gamma$ varies somewhat with the thickness of the sample, but is in the neighborhood of $\Gamma~\sim 17\,\gamma$. The semiquantitative agreement with collision theory is intriguing, but our atoms do not collide at all. More likely, the effective linewidth $\Gamma$ arises from the linewidths of the cooperative radiation modes~\cite{RUS96,JEN12a} in the sample.

The inhomogeneous broadening apparently emphasizes mean-field physics at the expense of correlations between adjacent atoms. To demonstrate how it works, we sketch a formal solution to the analog of Eqs.~\eq{LINEQS} for two atoms 1 and 2 with different resonance frequencies, hence different polarizabilities $\alpha_1$ and $\alpha_2$. 
Averaging over the corresponding atomic positions would then, in principle, allow calculation of the transmitted light through this idealized two-atom `sample', analogously to our numerical studies.
Evaluating the field on, say, atom 2 that is generated by the incident field and the light scattered from atom 1 yields
\bea
&&\bE(\br_2) = (1-\alpha_1\alpha_2 \G\G)^{-1}[\cbE_0(\br_2) + \alpha_1\G\cbE_0(\br_1)]\nonumber\\
&&=\cbE_0(\br_2)  + \alpha_1\G\cbE_0(\br_1) + \alpha_1\alpha_2 \G\G\cbE_0(\br_2) + \ldots\,.
\eea
The second line shows the beginning of the expansion of the inverse of the operator $(1-\alpha_1\alpha_2 \G\G)$, with $\G\equiv\G(\br_1-\br_2)=\G(\br_2-\br_1)$. The first term is the free field on atom 2; in the second term the free field excites atom 1, which sends its dipolar field back on atom 2; in the third term the free field excites atom 2, which sends a dipolar field to excite atom 1, which sends a dipolar field back on atom 2. Further terms in the expansion come out the same way reflecting repeated photon exchanges between the atoms. Such {\em recurrent} scattering processes in which (here) a classical wave scatters more than once by the same atom are responsible for the cooperative phenomena and the emergence of subradiant and superradiant resonances~\cite{JAV97,Jenkinsoptlattice,JEN12a}.

Let us now regard atom 2 as the spectator and imagine averaging over the position of atom 1. This operation faces major mathematical obstacles because of the divergence of $\G(\br_1-\br_2)$, but we do not attempt to sort them out because these problems are evidently similar for homogeneously and inhomogeneously broadened samples. Upon averaging, the second term becomes the mean-field contribution radiated by an assumedly continuous polarization, and further terms represent repeated photon exchanges between the atoms. Next add the inhomogeneous broadening $\omega_D$. To the order of magnitude, averaging over the resonant frequencies suppresses the polarizability by a factor of $\gamma/\omega_D$. Thus, the first nontrivial term in the expansion corresponding the mean-field polarization gets suppressed by this small factor, and the higher terms by higher powers of the small quantity $\gamma/\omega_D$. Qualitatively, repeated photon exchanges are de-emphasized because in such processes both the emitter and the absorber are off resonance. 

Analogously, one would expect that in a many-atom sample the transition from homogeneously broadened to inhomogeneously broadened phenomenology  takes place when the inhomogeneous broadening $\omega_D$ and the effective linewidth $\Gamma$ are comparable. This is, in fact, what we observe in the numerical experiments.

Why mean-field theory worked for the M\"ossbauer experiment~\cite{ROH10} even though the sample was homogeneously broadened is also easy to understand qualitatively. There the nuclei were effectively in a cavity that  directed the radiation repeatedly back on the nuclei. This clearly de-emphasizes the correlations between nearby radiators in favor of the mean field.

From classical-electrodynamics simulations, we have found qualitative features in the optical response of a homogeneously broadened (ultralow-temperature) dense atomic sample that are at variance with the time-honored pictures of local-field corrections and collective Lamb shifts. However, an inhomogeneous broadening (random distribution of atomic resonance frequencies) restores the agreement with the traditional theory. It turns out that the established picture is a mean-field approximation that breaks down when dipole-dipole interactions set up correlations between the radiators. Inhomogeneous broadening or a cavity that collects the radiation and redirects it back on the radiators both de-emphasize the correlations, and nudge the physics toward the mean-field theory. These observations are likely to be relevant when the experiments move toward dense low-temperature samples, for instance to improve the quantum nature of the atom-field coupling for the benefit of quantum metrology or possibly quantum information processing.

We acknowledge support from NSF, Grant No. PHY-0967644, and EPSRC. Most of the computations were done on Open Science Grid, VO Gluex, and on the University of Southampton High Performance Computing facility Iridis 4.

\end{document}